\documentclass[journal=jpca,manuscript=article]{achemso}
\usepackage{booktabs}
\usepackage{titlesec}
\usepackage{siunitx}
\usepackage[version=3]{mhchem} 
\usepackage{physics}
\usepackage{subcaption}

\author{Sarai Dery Folkestad}
\affiliation{Department of Chemistry, The Norwegian University of Science and Technology, Trondheim, Norway}
\email{sarai.d.folkestad@ntnu.no}
\author{Ida-Marie Høyvik}
\affiliation{Department of Chemistry, The Norwegian University of Science and Technology, Trondheim, Norway}
\email{ida-marie.hoyvik@ntnu.no}

\title{An orthogonal electronic state view on charge delocalization and transfer.}

\begin{document}


\begin{abstract}
We present a configuration interaction (CI) framework which serves quantitative and conceptual purposes for charge delocalization and electron transfer processes in molecular systems.  The electronic Hamiltonian is expressed in a basis of charge-localized determinants and used to independently generate adiabatic CI states and charge-localized CI states, both of which are unambiguously defined through a diagonalization procedure. The CI framework offers a simple interpretation of adiabatic states as resonance hybrids of different electron distributions, providing a simple picture for discussing charge-delocalization in chemical bonding. The charge-localized states serve as a convenient orthogonal representation of initial and final states in electron transfer processes, 
and provides an unambiguous definition of their electronic coupling. These 
two models enable an analysis of the water dimer hydrogen bond. We demonstrate that although the overall charge delocalization is small, the occurrence of particular ionic contributions are crucial to get the correct electronic description. 
\end{abstract}

Electronic-structure theory should, ideally,  offer conceptual simplicity for interpreting and extracting information of integer electron transfer or delocalization of electronic charge in chemical bonding. However, standard formulations of the electronic wave function blurs out this information and \textit{ad hoc} measures, such as population analysis and energy decomposition based schemes, are needed to recover the information. In this letter, we introduce a configuration interaction (CI) framework for the electronic wave function which contain chemical concepts related to bonding, while providing quantities central to understanding and quantifying integer and partial electron transfer (charge delocalization). The framework offers a clear and simple interpretation of adiabatic states\cite{Smith:1969aa} as the resonance hybrids\cite{pauling_book} between determinants of different electron distributions within a molecular system. Categorizing determinants in terms of electron distributions further enables a conceptually straightforward definition of charge-localized states, useful for computing initial and final states of integer electron transfer processes,\cite{Mikkelsen:1987aa,Subotnik:2009aa} and their Hamiltonian matrix elements (electronic couplings).  

 Electronic coupling elements have significance for several important fields, \cite{Marcus:1956aa,barbara_contemporary_1996,voorhis_review,Hsu:2009aa,Futera:2017aa,nitzan:2001,Nitzan:2001aa,Valeev:2006aa,Naaman:2022aa} and the use of electronic couplings to predict electron transfer rates through e.g., Marcus theory\cite{Marcus:1956aa} has resulted in a large interest in computing such couplings for many decades.\cite{Marcus:1985aa,Newton:1991aa,Braga:1993aa,cave_generalization_1996,cave_calculation_1997,Hsu:1997aa,Pavanello:2011aa,Migliore:2009aa,Atchity:1997aa,Ruedenberg:1993aa,nakamura_direct_2001,nakamura_extension_2003,Cembran:2009aa,futera_electronic_2017,subotnik_constructing_2008,Kondov:2007aa,Pavanello:2013aa,gray_long-range_2005,Hsu:2009aa,Lin:2018aa,pourtois_photoinduced_2002,song_construction_2008,Grofe:2017aa,Biancardi:2017aa,Storm:2019aa,Rikus:2025aa,Illesova:2025aa} The initial and final electronic states of the electron transfer process are in the literature usually referred to as diabatic states, which are required to fulfill some chosen criteria, for example by designing diabatic states with desirable characteristics\cite{Atchity:1997aa,Ruedenberg:1993aa,accomasso_diabatization_2019,pacher_approximately_1988,nakamura_direct_2001,Hiberty:2002aa,wu_extracting_2006,oberhofer_electronic_2010,Pavanello:2011aa,Grofe:2017aa} or invoking some physical observable such as the dipole operator\cite{cave_calculation_1997,cave_generalization_1996}.
 However, we note that diabatic states are formally states in which the nuclear derivative coupling is zero (or small),\cite{Smith:1969aa,baer_adiabatic_1975,OMalley:1971aa} although  these do not in general exist~\cite{mead_conditions_1982}.  For that reason we avoid this terminology and rather use the term charge-localized states, as also used by others.~\cite{Pavanello:2011aa} 
 
 We may summarize the advantages of our framework in three points; (1) the adiabatic states themselves directly contain information on the nature of the charge delocalization, providing qualitative and quantitative insights to the role of charge delocalization in chemical bonding, (2) adiabatic states and charge-localized states  can be generated independently from the same Hamiltonian matrix representation, (3) all charge-localized ground and excited states are orthogonal to each other and generated by a well-defined diagonalization procedure.

We consider a molecular electronic Hamiltonian,
\begin{equation}
\label{h_2nd}
H=\sum_{PQ} h_{PQ}a_P^\dagger a_Q+\frac{1}{2}\sum_{PQRS}g_{PQRS}a_P^\dagger a_R^\dagger a_S a_Q+h_\mathrm{nuc}
\end{equation}
expressed using spin-orbitals in the second quantization formalism,\cite{helgaker2013molecular}
$h_{PQ}$ and $g_{PQRS}$ are one- and two-electron integrals
in the spin-orbital basis, and $h_{\mathrm{nuc}}$ is the nuclear repulsion energy.

We consider a molecular system consisting of two non-covalently bonded molecules, named, for simplicity, subsystems $A$ and $B$.
The occupied and virtual spin-orbitals $\{\phi_P\}=\{\phi_I,\phi_J, \dots, \phi_A,\phi_B,\dots\}$ for the composite molecular system are spatially localized such that each local occupied and each local virtual spin-orbital may be assigned to either $A$ or $B$.  We denote spin-orbitals assigned to $A$  with unbarred indices, $\{\phi_p\}=\{\phi_i,\phi_j, \dots, \phi_a,\phi_b, \dots\}$, and spin-orbitals assigned to $B$ with barred indices $\{\phi_{\bar p}\}=\{\phi_{\bar i},\phi_{\bar j}, \dots, \phi_{\bar a},\phi_{\bar b}, \dots\}$. We emphasize that $\{\phi_p\}\cup\{\phi_{\bar p}\}$ forms an orthonormal basis for the composite system. 
The last decade has seen an advancement in optimization algorithms\cite{Hoyvik:2012ab} and localization functions\cite{Jansik:2011aa,Hoyvik:2012aa} which can generate spatially localized occupied and virtual orbitals.\cite{hoyvik_characterization_2016} Historical localization functions such as Pipek-Mezey\cite{pipek_fast_1989}, Edmiston-Ruedenberg\cite{edmiston_localized_1963,edmiston_localized_1965} and Foster-Boys\cite{foster_canonical_1960,edmiston_localized_1963} usually adequately localize occupied orbitals, while the locality of the virtual orbitals are dependent on the molecular system and the chosen atomic orbital basis set. 
The usefulness of the local orbitals depend on what they are intended for, and it can be seen that localized orbitals with a complicated nodal structure (as generated by e.g., powers of the variance and fourth-moment localizations) may be ineffective for e.g., local correlation models.\cite{Hansen:2020aa} 
For the work in this letter, the important point is that the orbitals used are local to one of the subsystems, and that orbital tails are primarily due to the mathematical requirement of orthogonalization. For a discussion  on orthogonalization tails, see the \textit{Supporting Information}.

A reference determinant (the Hartree-Fock determinant in case of local Hartree--Fock orbitals), in the basis $\{\phi_p\}\cup\{\phi_{\bar p}\}$, may be written as
\begin{equation}
\label{eq:reference_det}
|\Phi\rangle= \prod_{i=1}^{N_A} a_i^\dagger \prod_{\bar i=1}^{N_B} a_{\bar i}^\dagger \ket{\mathrm{vac}},
\end{equation}
where $N_A$ and $N_B$ is the number of electrons on subsystem $A$ and $B$, respectively, in the reference (denoted the reference electron distribution). The total number of electrons in the composite system is $N = N_A + N_B$. 

Excited determinants are defined by moving electrons from occupied spin-orbitals to virtual spin-orbitals, as is standard practice. However, in the local spin-orbital basis we  may categorize the determinants based on their electron distributions relative to the reference determinant; every determinant can be labeled with an integer $\lambda$, which denotes the number 
of electrons moved between subsystems relative to the reference determinant. 
We therefore introduce the notation
\begin{equation}
\ket{I^\lambda}: \quad \text{determinant }I\text { with electron distribution } (N_A+\lambda,N_B-\lambda),
\end{equation}
with $\lambda = 0, \pm 1, \pm2, \ldots$ We have chosen a convention where subsystem $A$ has received $\lambda$ electrons from  $B$, relative to the reference. The charge-localized determinants $\{\ket{I^\lambda}\}$ form an orthonormal $N$-electron basis, i.e., $\bra{I^\lambda}\ket{J^\tau}= \delta_{IJ}\delta_{\lambda\tau}$. 

\begin{figure}[H]
    \centering
    \includegraphics[width=0.8\linewidth]{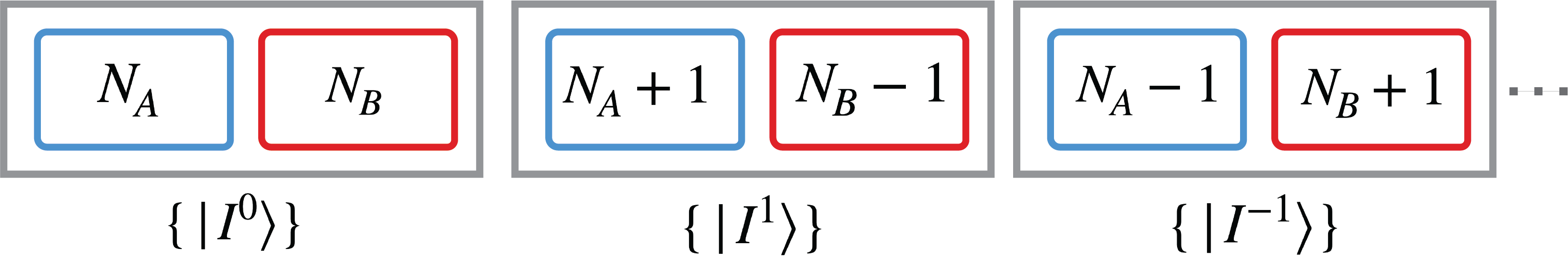}
    \caption{The charge-localized determinants, organized in terms of their eigenvalues of $\hat{n}_A$ and $\hat{n}_B$ ( $N_A + \lambda$ and $N_B - \lambda$, respectively). The total number of electrons in all determinants is $N = N_A + N_B$. The union of all determinants spans the N-electron space for the chosen CI truncation level.}
    \label{fig:determinants}
\end{figure}

The determinants $\{\ket{I^\lambda}\}$ are eigenstates of the total number operators, $\hat{n}=\hat{n}_A+\hat{n}_B$ with eigenvalue $N$ (the total number of electrons). We note that the total number operator naturally partitions into a number operator for $A$ and a number operator for $B$ when expressed in the local spin-orbital basis. The determinants $\{\ket{I^\lambda}\}$ are also eigenstates of  $\hat{n}_A$ and $\hat{n}_B$, where the eigenvalues depend on $\lambda$, see Figure \ref{fig:determinants}.

\begin{figure}[H]
    \centering
    \includegraphics[width=0.8\linewidth]{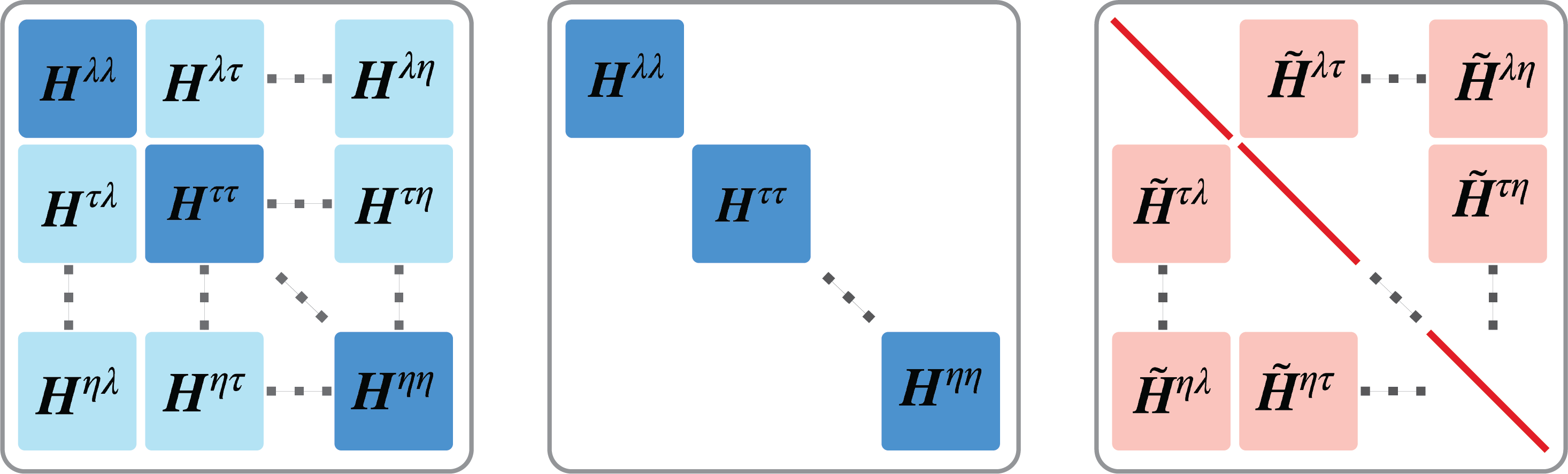}
    \caption{\textbf{Left:} The molecular electronic Hamiltonian in the charge-localized $N$-electron determinant basis. \textbf{Middle:} The approximate block-diagonal Hamiltonian. \textbf{Right:} the molecular electronic Hamiltonian in the basis of eigenvectors of the approximate block-diagonal Hamiltonian.}
    \label{fig:hamiltonians}
\end{figure}

The Hamiltonian of eq. \eqref{h_2nd} can be represented in the charge-localized $N$-electron determinant basis $\{\ket{I^\lambda}\}$,
\begin{equation}
H_{I^\lambda J^\tau} = \bra{I^\lambda} \hat{H} \ket{J^\tau},
\end{equation}
and if the determinant basis is sorted according to electron distribution, we obtain the block structure illustrated in Figure \ref{fig:hamiltonians} (left). Variational minimization of the energy with respect to the expansion coefficients yields the eigenvalue equation, 
\begin{equation}
\label{eq:standard_fci}
\mathbf{H} \mathbf{C}_k = E_k \mathbf{C}_k,
\end{equation}
and the (adiabatic) CI wave function is, in terms of the charge-localized determinants, given by
\begin{equation}
\label{eq:ci_state_cl_basis}
\ket{\Psi_k} = \sum_\lambda \sum_{I^{\lambda}} C_k^{I^\lambda}\ket{I^\lambda}.
\end{equation}

Performing the CI procedure in a charge-localized determinant basis makes it possible to determine the expected number of electrons on a subsystem in a given electronic state. For example, the average number of electrons on subsystem $A$ is given by
\begin{equation}
\label{eq:expected_na}
\langle \hat{n}_A \rangle_k = \bra{\Psi_k}\hat{n}_A\ket{\Psi_k} = \sum_\lambda\sum_{I^\lambda} |C^{I^\lambda}_k|^2(N_A+\lambda).
\end{equation}
Furthermore, by summing only over a subset of electron distributions one may quantify how important this process is in the CI state. For example, the importance of moving $\lambda$ electrons from B to A in the CI state is given by
\begin{equation}
\label{eq:prob_lambda}
P_k^\lambda=\sum_{I^\lambda} |C^{I^\lambda}_k|^2.
\end{equation}
We use the symbol $P$ to emphasize that due to the normalization of the CI states the quantity in eq. \eqref{eq:prob_lambda} can be considered as a probability for the process defined by $\lambda$ within the CI state.

Rather than carrying out a diagonalization of the full Hamiltonian matrix, as done in standard CI, we may diagonalize in the subspaces of each of the electron distributions, i.e., within the blocks $\mathbf{H}^{\lambda \lambda}$,
\begin{equation}
\label{eq:cl_ci}
\mathbf{H}^{\lambda \lambda} \tilde{\mathbf{C}}_k^\lambda = \tilde{E}_k^\lambda  \tilde{\mathbf{C}}_k^\lambda
\end{equation}
where $\tilde{E}_k^\lambda$ is the energy of the $k$th charge-localized CI state with electron  distribution $\lambda$. The charge-localized FCI wave function for state $k$ is,
\begin{equation}
\label{eq:cl_ci_state}
\ket{\tilde{\Psi}_k^\lambda} = \sum_{I^\lambda} \tilde{C}_k^{I,\lambda} \ket{I^\lambda},\quad\quad \bra{\tilde{\Psi}_k^\lambda}\ket{\tilde{\Psi}_l^\tau} = \delta_{kl}\delta_{\lambda\tau}.
\end{equation}
Solving eq. \eqref{eq:cl_ci} for each electron distribution amounts to diagonalizing the (approximate) Hamiltonian matrix illustrated in Figure \ref{fig:hamiltonians} (middle). Importantly, the charge-localized states are orthonormal within an electron distribution and orthogonal between different electron distributions, since they represent different eigenvectors of a Hermitian matrix.
By transforming the full Hamiltonian matrix to the basis of charge-localized states, Figure \ref{fig:hamiltonians} (right), we identify the electronic coupling elements ($\tilde{\boldsymbol{H}}^{\lambda\tau}$) between different states of the different charge distributions $\lambda$ and $\tau$. We emphasize that this procedure gives an unambiguous definition of mutually orthogonal ground and excited charge-localized states, and the electronic coupling elements between charge-localized states of different charge distributions.

Details on the implementation can be found in the \textit{Supporting Information}, and we first illustrate the information available in the adiabatic CI states, before proceeding to show results for the charge-localized states  and show how a combination of both approaches can be used for investigation of chemical bonding.  We present results for (H$_2$)$_2^+$\cite{barbara_contemporary_1996,schatz_ratner} (non-bonded), He$_2^+$ (covalent bond) and (H$_2$O)$_2$ (hydrogen bond), and we note that throughout the results,  curves plotted in red are results of the adiabatic CI calculations while blue curves denote results from charge-localized states.

\begin{figure}
     \centering
     \includegraphics[width=0.8\linewidth]{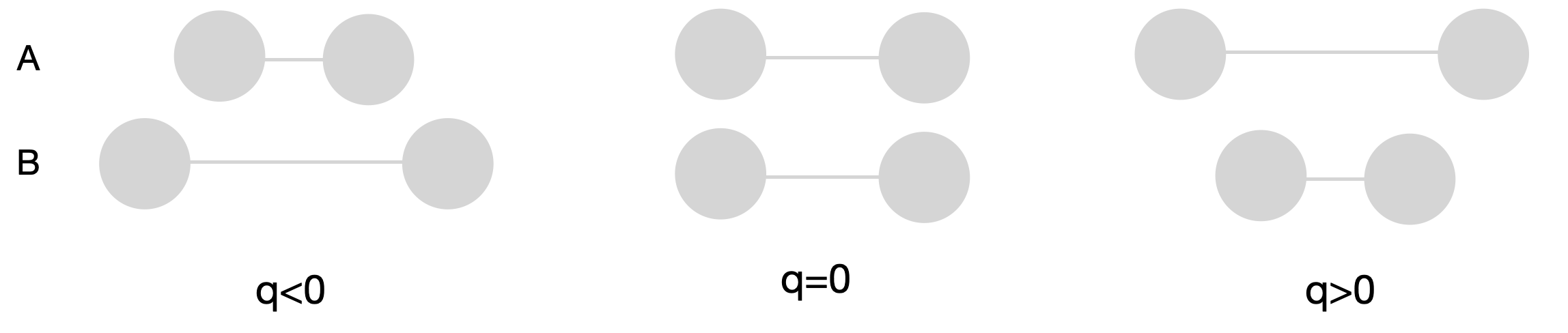}
     \includegraphics[width=\linewidth]{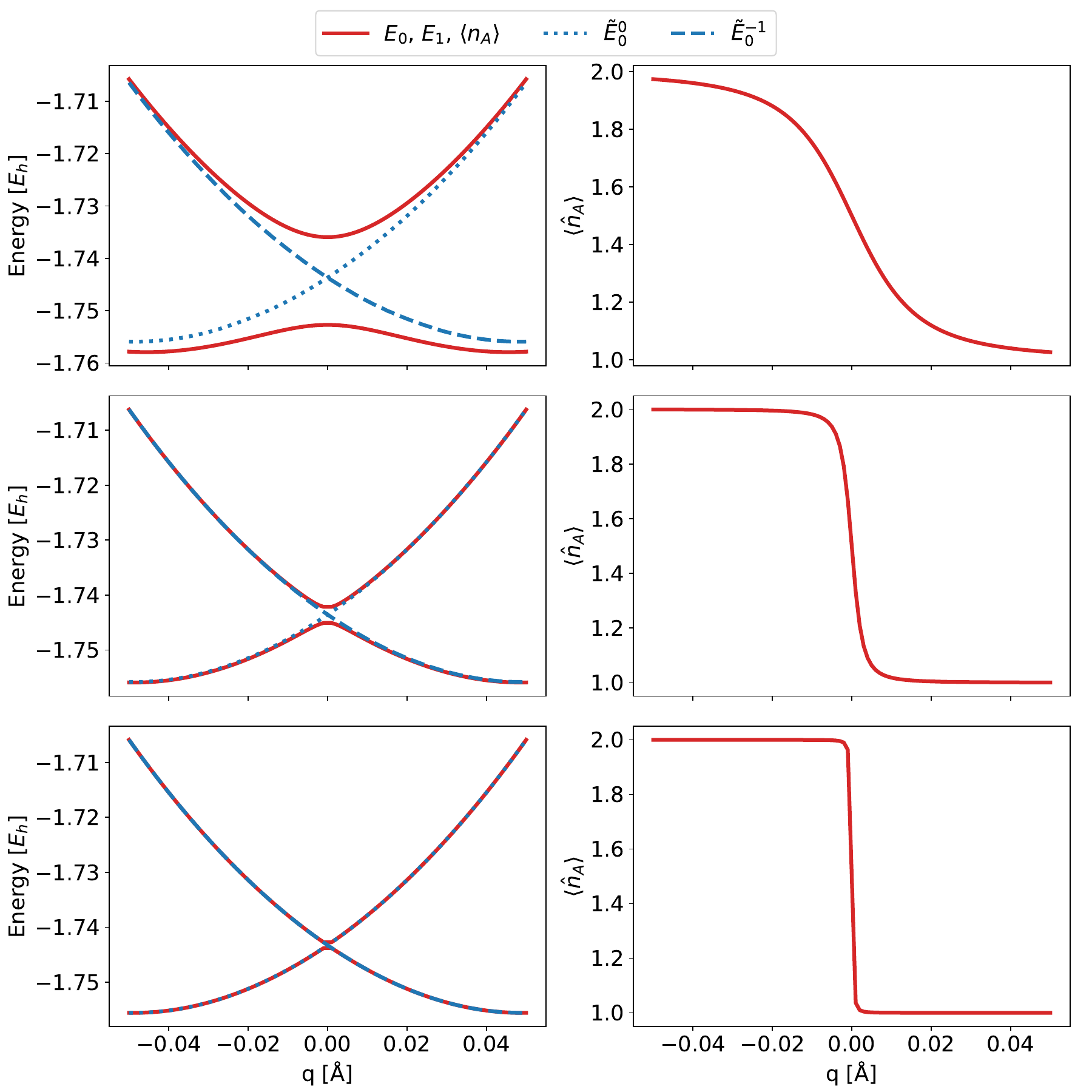}
     \caption{(H$_2$)$_2^+$ with  $\SI{3}{\angstrom}$ (top row),  $\SI{4}{\angstrom}$ (middle row) and $\SI{5}{\angstrom}$ (bottom row) separation between monomers. The plots contain the potential energy curves for ground and excited FCI and ground states of charge-localized states for $\lambda=0$ and $\lambda=-1$ (left column) and the expected charge on monomer $A$ (right column). The coupling element is approximately constant along the reaction coordinate, but it varies depending on the separation between the H$_2$ monomers: $|\tilde{H}^{0-1}_{00}| = \SI{0.227}{\eV}$ at $\SI{3}{\angstrom}$, $|\tilde{H}^{0-1}_{00}| = \SI{0.038}{\eV}$ at $\SI{4}{\angstrom}$, and $|\tilde{H}^{0-1}_{00}| = \SI{0.005}{\eV}$ at $\SI{5}{\angstrom}$. All results are generated using cc-pVDZ.}
     \label{fig:h22plus_placeholder}
 \end{figure}

The red curves in Figure \ref{fig:h22plus_placeholder} represent results from the FCI calculation of (H$_2$)$_2^+$ as a function of the  bond length difference between the monomers ($q$) for inter-monomer distances of $\SI{3}{\angstrom}$ (top row),  $\SI{4}{\angstrom}$ (middle row) and $\SI{5}{\angstrom}$ (bottom row).  The left column of Figure \ref{fig:h22plus_placeholder} contains the potential energy surfaces, and the right column contains the expected charge on monomer $A$ for the ground state, as computed using eq. \eqref{eq:expected_na}. At $\SI{3}{\angstrom}$ distance between the monomers (top row), a large splitting of the ground ($E_0$) and first excited ($E_1$) can be seen.  The expected number of electrons on $A$, see Figure \ref{fig:h22plus_placeholder} (top, right), goes smoothly from two electrons ($q\ll0$) to one electron ($q\gg 0$) through a relatively wide range around $q=0$.  I.e., in this region, parts of the electronic density is shared between the two monomers. 

Charge transfer processes is sometimes discussed in terms of electronic timescales, see e.g., Ref. \citenum{may_charge_2011}, p. 315-318. For large electronic coupling elements, such as in (H$_2$)$_2^+$ at $\SI{3}{\angstrom}$, the electron transfer process is said to be adiabatic. As seen from the wave function, the probability distribution for the electrons is smeared between the systems, i.e., it is as a resonance hybrid between two distinct electron distributions. 
Such a resonance is, naturally, most notable for a system with a covalent bond, such as  He$_2^+$.  The expected number of electrons for both subsystems in  He$_2^+$ (helium atoms) is computed to be 1.5 for FCI. I.e., He$_2^+$  represents one limiting case where the subsystems equally share the electrons.
The other limiting case, when no charge is shared and charge transfer is said to be non-adiabatic, happens for small coupling elements (for instance for (H$_2$)$_2^+$ at $\SI{5}{\angstrom}$). 
The magnitude of the coupling element is a necessary, but not a sufficient condition for resonances between different charge distributions: the energies of the charge-localized states must also be relatively close, enabling a lowering of the energy upon e.g., variational optimization.

The transition between delocalized charge and localized charge can be seen when comparing results for (H$_2$)$_2^+$ at $\SI{3}{\angstrom}$, $\SI{4}{\angstrom}$ and $\SI{5}{\angstrom}$ distance between subsystems (Figure \ref{fig:h22plus_placeholder} top, middle and bottom, respectively). At short subsystem separations,  (H$_2$)$_2^+$ has partial bonding character in the region around $q=0$ as indicated by the delocalization of charge and the energy lowering relative to the same value of $q$ for longer inter-monomer distances. For longer subsystem separations, the  energetic separation of the adiabatic ground and excited states is smaller, and the expected number of electrons on monomer $A$ goes toward a step function. I.e., for longer subsystem separations, there is no region of delocalized charge, and only integer electron transfer occurs. 

  \begin{figure}[H]
     \centering
     \includegraphics[width=\linewidth]{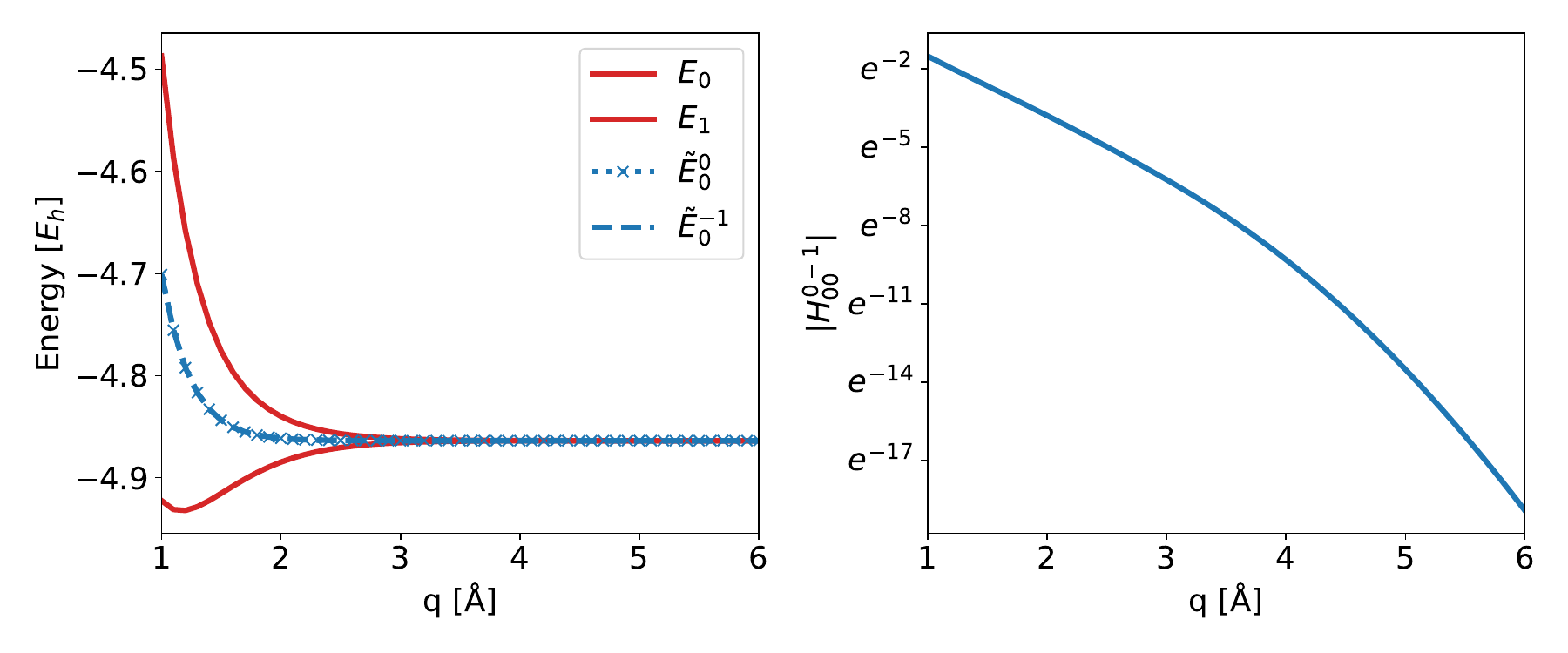}
     \caption{Left: The ground ($E_0$) and first excited ($E_1$) FCI states and the charge-localized ground states for He He$^+$ ($\tilde{E}_0^0$) and  for He$^+$  He ($\tilde{E}_0^{-1}$)  for He$_2^+$. Right: The electronic coupling between the charge-localized ground states for He$^+$ ($\lambda=0$) and He$^+$ He ($\lambda=-1$). All results are generated using 6-31G$^*$.}
     \label{fig:He2+}
 \end{figure}

We now discuss results generated for charge-localized CI states. The ground state charge-localized energy curves for electron distributions $\lambda=0$ ($\tilde{E}_0^{0}$) and $\lambda =-1$ ($\tilde{E}_0^{-1}$) is plotted in Figures \ref{fig:h22plus_placeholder} and \ref{fig:He2+}. We first look at Figure \ref{fig:h22plus_placeholder}, where the charge-localized FCI ground state for $\lambda=0$  (H$_2$ H$_2^+$) and ground state for  $\lambda=-1$ (H$_2^+$ H$_2$) is plotted for (H$_2$)$_2^+$.  The electronic coupling element between these charge-localized states are computed to be  $|\tilde{H}^{0-1}_{00}| = \SI{0.227}{\eV}$ at $\SI{3}{\angstrom}$, $|\tilde{H}^{0-1}_{00}| = \SI{0.038}{\eV}$ at $\SI{4}{\angstrom}$, and $|\tilde{H}^{0-1}_{00}| = \SI{0.005}{\eV}$ at $\SI{5}{\angstrom}$. The electronic coupling elements are computed at each $q$, but they are found to be constant across the chosen reaction coordinate to within decimal points given here. At $\SI{3}{\angstrom}$, there is a strong coupling between the charge-localized states, and this can also be seen from the fact that the adiabatic energy curves (E$_0$ and E$_1$) deviate from the charge-localized energy curves. At $\SI{5}{\angstrom}$, the electronic coupling is weak and the charge-localized curves are superimposed on the adiabatic curves.  The energy splitting between the adiabatic states $E_0$ and $E_1$ is seen to reflect 2$|\tilde{H}_{00}^{0-1}|$, which it would be in a two-state calculation (see e.g. Ref. \citenum{may_charge_2011}, p. 41). The charge-localized CI ground states for $\lambda=0$ (He He$^+$) and $\lambda=-1$ (He$^+$ He) are given in Figure \ref{fig:He2+} (left), and they are degenerate since the electron distributions  $\lambda=0$ and $\lambda=-1$ are equivalent.  The electronic coupling between them, Figure \ref{fig:He2+} (right), decays exponentially with internuclear distance, as seen from the near-linear form on the base $e$ logarithmic scale. Hence, the features of the energies and electronic coupling elements computed using charge-localized states is consistent with the use of so-called diabatic states in the literature. To make a quantitative comparison, we compare electronic coupling elements for internuclear distances 2 Å and  $2\sqrt{2}$ Å for  He$_2^+$  to results from  Ref. \citenum{subotnik_constructing_2008}.  This is presented in Table \ref{tab:he2_plus}. As is seen from Table \ref{tab:he2_plus}, the results for the coupling elements in the charge-localized CI states are similar the results produced by the Boys localized states. For further comparisons with other reported methods, see the \textit{Supporting Information}.

\begin{table}[H]
    \centering
    \begin{tabular}{ c c c c}
    \toprule
    $R $ & $\tilde{H}_{00}^{0-1}$ (FCI)  & $\tilde{H}_{00}^{0-1}$ (CISD) &  $H_{AB}$ (Ref.\citenum{subotnik_constructing_2008})\\
    \midrule
      $2$ \si{\angstrom} &  $0.610$ eV &   $0.609$ eV & $0.617$ eV\\
       $2\sqrt{2}$ \si{\angstrom} & $0.082$ eV &  $0.082$ eV &$0.082$ eV\\    
      \bottomrule
    \end{tabular}
    \caption{Electronic coupling elements between He He$^+$ ($\lambda=0$) and He$^+$ He ($\lambda=-1$) computed using charge-localized versions of FCI and CISD using 6-31G$^*$.  Coupling elements, $H_{AB}$, for He$_2^+$ computed using complete active space self-consistent field  with three electrons distributed in four spin-orbitals and 6-31G$^*$ (equivalent to FCI) taken from Subotnik et al.\cite{subotnik_constructing_2008}.}
    \label{tab:he2_plus}
\end{table}

\begin{figure}[H]
    \centering
    \includegraphics[width=0.30\linewidth]{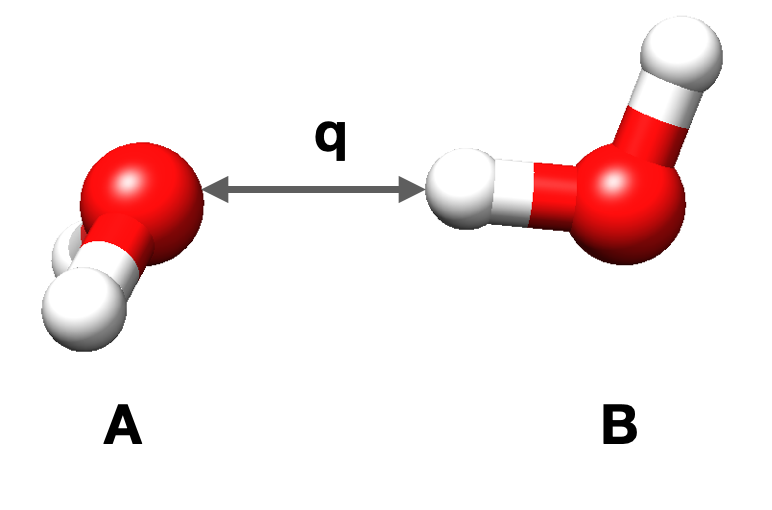}\\ 
    \includegraphics[width=0.48\linewidth]{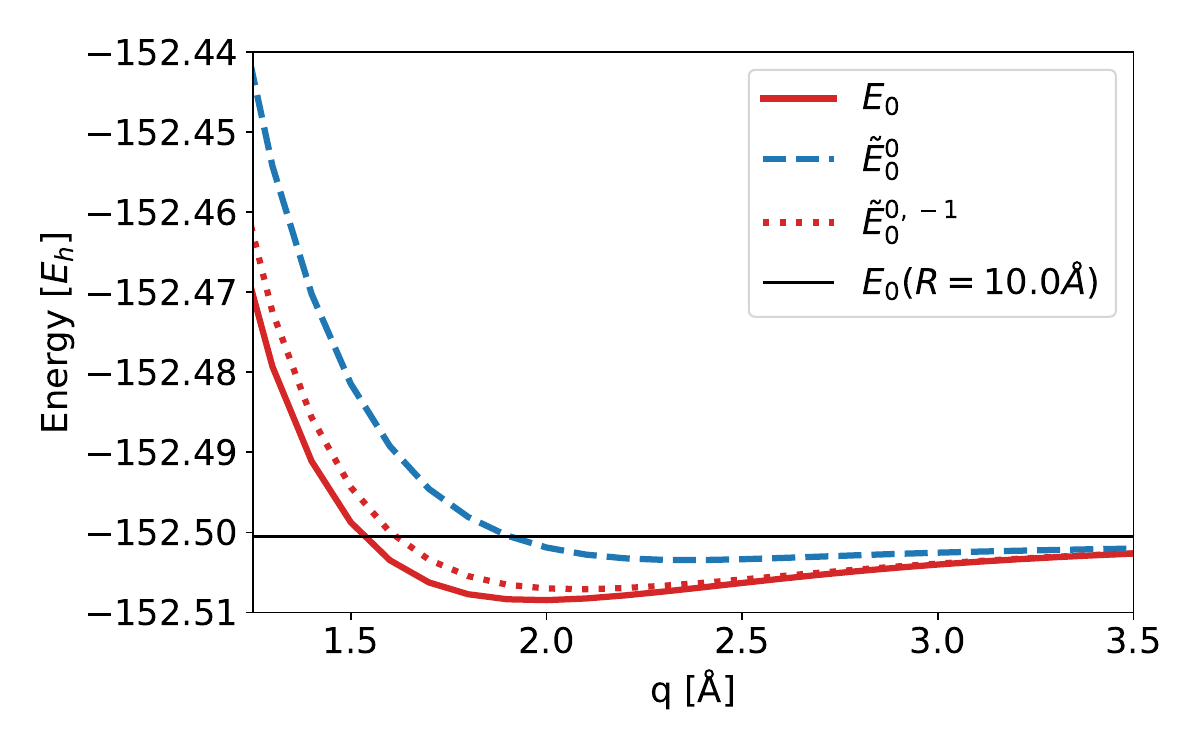} \includegraphics[width=0.48\linewidth]{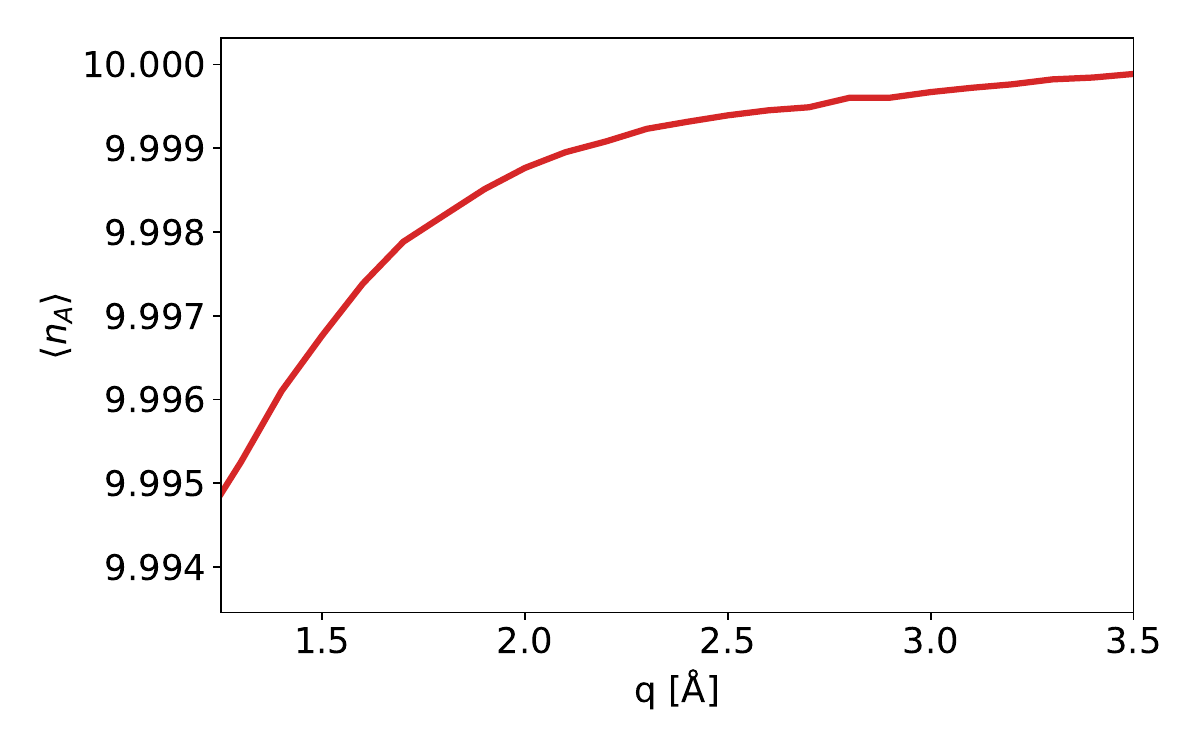}
    \caption{Left: CISD energy curve ($E_0$) and ground state charge-localized CISD energy curve for the ground state of the neutral-neutral electron distribution ($\tilde{E}_0^0$). The CISD ground state energy at a 10 Å separation is included. A restricted adiabatic CISD calculation which only involves $\lambda=0$ and $\lambda=-1$ is also included ($\tilde{E}_0^{0,-1}$). Right: The expected number of electrons on molecule $A$ (eq. \eqref{eq:expected_na}) for the CISD ground state as a function of $q$. At shorter $q$, water molecule $A$ exhibit a slight cationic character, implying that molecule $B$ exhibit a slight anionic character.
    All results are generated using aug-cc-pVDZ.}
    \label{fig:water_dimer}
\end{figure}

As seen above,  the charge-localized CI states gives us a framework consistent with that from electron transfer theory. We will now use results both from the adiabatic and charge-localized states in our framework to discuss the hydrogen bonding in the water dimer.  We note that the hydrogen bond acceptor (molecule $A$ in Figure \ref{fig:water_dimer}) is the donor of electronic density, whereas the hydrogen bond donor (molecule $A$ in Figure \ref{fig:water_dimer}) is the acceptor of electronic density. To avoid confusion, we will therefore simply refer to the molecules by using $A$ and $B$. The reaction coordinate $q$ is the distance between the hydrogen bonded oxygen and hydrogen. For a specification on the geometries, see the Ref.~\citenum{folkestad_2025_17058210}.  The reference electron distribution for this system is the neutral-neutral distribution, i.e., ten electrons in each water molecule. We note that the calculations presented here (CISD using an aug-cc-pVDZ basis set) are not intended to provide quantitative numbers for the non-covalent interaction energy between the water molecules, which would require a better (and size-extensive) $N$-electron model, an improved basis set and correction for basis set superposition error (BSSE). Rather, we use the water dimer as an illustrative example of  how concepts used introduced in this paper may offer insight to non-covalent interactions.

In Figure \ref{fig:water_dimer} we have plotted the CISD  ground state energy curve ($E_0$) for the water dimer as a function of $q$, together with the ground state charge-localized CISD energy curve for the neutral-neutral charge distribution ($\tilde{E}_0^0$). In addition, we have plotted an energy curve that is produced by allowing only electron distributions $\lambda =0$ and $\lambda = -1$ to mix.  Figure \ref{fig:water_dimer} (bottom left) shows that the CISD ground state energy curve ($E_0$) exhibits a  minimum, indicating the bonding interaction between the two water molecules. In Table \ref{tab:probabilities_adiabatic}  the total contribution from determinants of specific electron distributions (see Eq. \eqref{eq:prob_lambda}) to the CISD ground state is tabulated. For $q<5.0$ Å determinants with ionic electron distributions contribute. For example, at 1.$\SI{5}{\angstrom}$ the neutral-neutral determinants dominate ($P_0^0 = 0.9949$), but cationic-anionic ($P_0^{-1} = 0.0030$), anionic-cationic ($P_0^1 = 0.0009$) and doubly cationic-doubly anionic ($P_0^{-2} = 0.0004$) also contributes.  For values of  $q$ around the minimum of the CISD curve, we see that the cationic-anionic determinants are the most important determinants in addition to the dominating neutral-neutral determinants. For example, at $q=2.0$ Å, $P_0^{-1}$ is an order of magnitude larger than $P_0^{1}$  and $P_0^{-2}$.

\begin{table}[H]
    \centering
    \begin{tabular}{c c c c c c }
    \toprule
    $q$ & $1.5\;\si{\angstrom}$ & $2.0\;\si{\angstrom}$ &  $3.0\;\si{\angstrom}$ & $4.0\;\si{\angstrom}$ & $5.0\;\si{\angstrom}$  \\
    \midrule 
     $P_0^{-2}$ & 0.0004 & 0.0001 & 0.0000 & 0.0000 &  0.0000 \\
     $P_0^{-1}$ & 0.0030 & 0.0019 & 0.0006 & 0.0001 &  0.0000 \\
     $P_0^{0}$  & 0.9949 & 0.9977 & 0.9994 & 0.9999 &  1.0000 \\
     $P_0^{1}$  & 0.0009 & 0.0003 & 0.0000 & 0.0000 &  0.0000 \\
    \bottomrule
    \end{tabular}
    \caption{Total probabilities, $P_0^\lambda$, for occurrence of cationic-anionic determinants ($P_0^{-1}$), doubly cationic-doubly anionic  determinants $P_0^{-2}$), neutral-neutral determinants ($P_0^{0}$) and anionic-cationic determinants (($P_0^{-1}$)) for the CISD ground state presented in Figure \ref{fig:water_dimer}}
    \label{tab:probabilities_adiabatic}
\end{table}

The qualitative and quantitative importance of the small occurrences of the ionic electron distributions can be seen from Figure \ref{fig:water_dimer} (bottom left), by comparing the CISD ground state energy $E_0$ to the charge-localized CISD ground state energy, $\tilde{E}_0^0$.   While the charge-localized CISD energy exhibits only a weak bonding interaction (as do Hartree-Fock for the water dimer) it is quantitatively and qualitatively different from the CISD ground state energy $E_0$ where the ionic configurations contribute. The presence of the cationic-anionic determinants in the CISD wave function is reflected in the expected number of electrons on molecule A, see Figure \ref{fig:water_dimer} (bottom right). At short $q$ the number of electrons on molecule $A$ is just below ten,  indicating a slightly cationic state of molecule $A$. The charge-transfer (or rather, charge delocalization) is on the order of millielectrons, with approximately 0.002 electrons transferred around the equilibrium bond length. This number is consistent with the numbers produced using DFT in combination with energy decomposition analysis\cite{Khaliullin:2007aa,Khaliullin:2008aa} based on absolutely localized molecular orbitals\cite{Nagata:2001aa,Khaliullin:2006aa}. As discussed in  Ref. \citenum{Khaliullin:2007aa}, charge-transfer on the millielectron scale is an order of magnitude smaller than charges computed using population analysis schemes, indicating that population analysis schemes overestimate the charge-delocalization. Our CISD results supports this claim.  The role of partial ionic character in hydrogen bonding has long been discussed in the literature\cite{coulson:1952,Bratoz:1967aa,pimentel:1971,Ratajczak:1973aa,Reed:1988aa}. However, as pointed out by Weinhold and Klein\cite{Weinhold:2012aa} as late as in 2012, most current textbooks describe hydrogen bonding with wording which only reflect the classical electrostatic picture (see discussion in Ref. \citenum{Weinhold:2012aa}).  Although there seem to be little controversy regarding that there is charge-transfer in hydrogen bonds, the amount is under debate.\cite{Stone:2017aa,Weinhold:2018aa}
The results in Figure \ref{fig:water_dimer} (bottom left) show that even if the ionic contributions are small (as seen from the millielectron charge-transfer),  they have a large effect on the wave function and energy. We therefore argue that the importance of ionic contributions in the wave function is not directly reflected in the amount of charge-transfer.

One may raise the question whether the ionic contributions in the water dimer calculation are finite basis set artifacts, i.e., whether they cause BSSE.
According to Schütz et al.,\cite{Schutz:1998aa} who considered interactions between monomers in the context of local correlation models, the doubly ionic contributions (double excitation from one monomer to the other) are responsible for the main portion of BSSE.
Whereas the charge-localized CISD model presented here 
only includes intra-monomer and exchange-dispersion excitations\cite{Schutz:1998aa,Bistoni:2020aa}, the adiabatic CISD state 
allows the doubly ionic contributions. 
From Table \ref{tab:probabilities_adiabatic} it can be seen that $P^{-2}_0$ is non-zero for small $q$. 
To evaluate the energetic effect of ignoring these doubly ionic contributions, we also present a calculation which omits determinants of all other electron distributions than the two dominant (neutral-neutral and cationic-anionic). The energy curve for this restricted CISD calculation (denoted by $\tilde{E}_0^{0,-1}$) is given in Figure \ref{fig:water_dimer}. By omitting the other types of determinants, the 
energy is higher compared to the full adiabatic calculation, as expected per the variational principle. The minimum is also shifted slightly to the right. However, the $\tilde{E}_0^{0,-1}$ still shows a significantly different curve than the charge-localized state of neutral-neutral character ($\tilde{E}_0^0$). 
Hence, there is reason to believe that the presence of the cationic-anionic states are not 
an artifact of using a finite basis, and that charge delocalization---not only electrostatic interactions---is central to a quantitative description the hydrogen bond in the water dimer.

In this paper we have introduced a set of $N$-electron orthonormal determinant basis constructed from a common and localized orbital space for interacting subsystems.  Each determinant can be categorized  according to its electron distributions across the subsystems, and it represent a valid determinant (obeying the Pauli principle) for the composite system. 
The charge-localized determinant basis provides a powerful framework for the CI expansion. From the same electronic Hamiltonian matrix representation, we may independently generate standard (adiabatic) CI states or charge-localized states. Standard CI states are  computed by diagonalizing the full electronic Hamiltonian, while  diagonalizing within subspaces of specific electron distributions gives rise to charge-localized CI states. The charge-localized CI ground and excited states are orthonormal states with specific electron distributions, and they are therefore suitable for representing  initial and final states of e.g. electron transfer processes. In the charge-localized basis, the off-diagonal elements of the electronic Hamiltonian gives the electronic coupling between the different charge-localized ground and/or excited states. Furthermore, since the standard CI states are expressed in the charge-localized determinant basis, we can get charge-transfer and charge delocalization information directly from the CI wave function. The presented CI framework unifies illustrative chemical concepts such as resonances from valence bond theory with how correlated electronic wave function models are constructed. As en example we have presented results for the water dimer, showing that the occurrence of a particular type of ionic determinants is crucial for the wave function despite charge delocalization effects being small (millielectron scale). The resonances between different electron configurations further provide a conceptually simple framework for understanding and discussing how an integer electron transfer process occurs, while providing necessary quantities such as charge-localized ground and excited states and their electronic coupling elements.

\section*{Supporting Information}
The \textit{Supporting Information} for this letter contains an extended review of the literature, implementation and computational details, a discussion about orthogonalization tails in orbital localization, and additional data considering the basis set and localization function dependence and comparisons to other approaches for the electronic coupling from the literature. Molecular geometries are provided in Ref.~\citenum{folkestad_2025_17058210}.

\section*{Author Contributions}
Both authors has contributed equally.

\section*{Acknowledgments}
I-M.H. acknowledge funding from the Research Council of Norway through FRINATEK project 325574 and support from the Centre for Advanced Study in Oslo, Norway, which funded and hosted her Young CAS Fellow research project during the academic year of 22/23 and 23/24.
We thank Jacob Pedersen for commenting on the manuscript and Bendik Støa Sannes for providing us with the water geometries used in the paper.

\bibliography{biblio_manager}

\end{document}